# High-Risk Memories? Comparative audit of the representation of Second World War atrocities in Ukraine by generative AI applications

Mykola Makhortykh (University of Bern), Victoria Vziatysheva (University of Bern), Maryna Sydorova (University of Bern/University of Fribourg)

**Abstract**: The rise of generative artificial intelligence (genAI) models poses new possibilities and risks for how the past is remembered by accelerating content production and altering the process of information discovery. The most critical risk is historical misrepresentation, which ranges from the distortion of facts and inaccurate depiction of specific groups to more subtle forms, such as the selective moralization of history. The dangers of misrepresentation of the past are particularly pronounced for high-risk memories, such as memories of past atrocities, which have a strong emotional load and are often instrumentalised by political actors. To understand how substantive this risk is, we empirically investigate how genAI applications deal with high-risk memories of the Second World War atrocities in Ukraine. This case is crucial due to the scope of the atrocities and the intense, often instrumentalised, contestation surrounding their memory. We audit the performance of three common genAI applications for different types of misrepresentation, including hallucinations and inconsistent moralization, and discuss the implications for future memory practices.

**Keywords**: moralization, memory, generative AI, hallucination, misrepresentation, Holocaust

## 1. Introduction

The emergence and growing accessibility of generative artificial intelligence (genAI) models, such as GPT or Llama, and their applications, such as ChatGPT or Gemini, have the potential to fundamentally transform how the past is remembered. GenAI accelerates the production of new content, including that related to historical events, in a multitude of formats, from text to image to video. It not only increases the volume of already abundant memory-related content in digital environments but also changes what information individuals discover about the past and how they do it. These transformations are evidenced by the growing adoption of genAI applications for teaching and learning about history (Morley 2024; Pope and Ma 2024), facilitating institution- and activist-based memory projects (Amichay 2023; Roth 2023), as well as, concerningly, for the distortion and appropriation of the past (Hoskins 2024; Menotti 2025), for political but also for commercial gains.

Social media platforms, such as Facebook or TikTok, are among those digital environments where the adoption of genAI for engaging with the past occurs particularly rapidly. It applies both to the content appearing on social media and the affordances these platforms provide to their users. The accessibility of genAI has resulted in artificially generated historical content flooding social media platforms, with examples ranging from fake historical photos showing victims of the Nazi concentration camps to videos portraying prominent historical events or offering brief overviews of specific communities' history. Simultaneously, social media platforms integrate genAI into their functionality; examples of such integration range from the deployment of AI assistants, such as Meta AI, which can answer user inquiries about a broad range of topics, including history, to the platform-based mechanisms of content transformation, for instance, animation of static images or video generation.

Despite the growing use of genAI to represent and interact with the past—and the consequent growth of scholarship on this topic—the long-term consequences of this technology for individual and collective remembrance remain unclear. The potential benefits include empowering the general public and researchers interested in learning about the past by enhancing their ability to process and interpret historical materials (Kansteiner 2022) and facilitating the production of new multimodal narratives about the past (Makhortykh et al. 2023a). Furthermore, genAI provides a possibility of tackling historical injustices by giving voice to genocide victims who did not manage to survive the atrocity and leave testimonies for future generations (Kozlovski and Makhortykh 2025). However, these benefits are contrasted by potentially severe environmental costs associated with genAI technologies (Merrill et al. 2025) and risks to cognitive memory mechanisms, including reduced capacity for information retention (Bai et al. 2023).

Besides the potential long-term negative effects, the use of genAI raises an immediate concern regarding its potential to misrepresent the past. Misrepresentation can take different forms. The irresponsible or adversarial application of genAI can distort historical facts, leading to misrepresentation of specific groups, such as people of colour being portrayed as Vikings or German soldiers (Milmo 2024). The same risk applies to individuals, for instance, in the case of AI-enabled digital duplicates whitewashing crimes of Nazi officials (Kozlovski 2025). However, there can also be other forms of misrepresentation that are more obscure and less recognised. One example is the moralization of the past, which positions genAI as a moral authority and nudges users towards specific normative interpretations of history (Presner et al 2026). While this may be advantageous under certain conditions, the selective production of moral judgements may be misleading and facilitate the instrumentalisation of the past.

The genAI-driven misrepresentations are of particular concern for the historical events that trigger strong emotional reactions and the interpretations of which are often contested (e.g. in the case of memories of mass atrocities). The combination of their societal significance and contestation also makes these memories particularly often instrumentalised, for instance, to justify agendas and policies promoted by populist and authoritarian actors (Shiller et al. 2023; Kalstein et al. 2024). Such *high-risk memories* are particularly threatened by the rise of genAI, which amplifies the risk of distortion and denial that, in turn, can challenge the memory-related ethical obligations within specific societies (e.g. to respect the memory of genocide victims) and facilitate the contestation and appropriation of the past. To assess the validity of these concerns, there is an urgent need for empirical research on how genAI represents and misrepresents high-risk memories, especially in a comparative perspective that accounts for the multiplicity of factors that affect genAI performance. Some of these factors include the prompt's language, the model choice, and the specific historical episode (Ulloa et al. 2025).

To address this need, we look at how genAI applications approach high-risk memories associated with Second World War atrocities in Ukraine. The case of Ukraine is particularly important in this context, not only due to the massive scope of atrocities committed against different population groups (e.g. Jews, Poles, and Ukrainians) who lived in the territories which are part of contemporary Ukraine, but also the intense contestation of memories associated with these atrocities both in Ukraine and neighbouring countries (for examples, see Narvselius 2012; Siddi 2017; Honcharenko 2025). Such a contestation often goes together with the instrumentalisation of memories about atrocities, for instance, as part of internal Ukrainian politics in the 2000s (Honcharenko 2025) or by the Kremlin as part of its propaganda campaigns aiming to justify Russian aggression against Ukraine (Griffin 2024).

The rest of the article is organised as follows: first, we discuss the risks of genAI, in particular text-focused models and applications, for misrepresentation of high-risk memories, and how these forms of misrepresentation can emerge. We then briefly introduce the research protocol we used to audit the performance of three genAI applications regarding high-risk memories in Ukraine and look at how these applications represent different aspects of mass atrocities from which different population groups in Ukraine suffered during the Second World War. We conclude with a discussion of the implications of our findings for history and memory.

## 2. What are the risks of text-generative AI for high-risk memories?

As a first step for empirically assessing the risks of genAI for high-risk memories, we need to consider what these risks can be and how they can emerge. While the topic of genAI-driven misrepresentation has attracted substantial scholarly interest (Gross 2023; Sun et al. 2024; Vázquez and Garrido-Merchán 2024), relatively few studies have explicitly examined how genAI misrepresents both the historical and recent past (Makhortykh et al. 2023b; Laba et al. 2025). While a few cases of misrepresentation, such as AI-generated images of mass atrocities, particularly the Holocaust, have attracted substantial public attention (Gkogkou 2025), there is still a limited understanding of what different forms of misrepresentation can be, especially in the case of high-risk memories.

A major challenge in studying the misrepresentation of the past by genAI is defining it. Both history and (collective) memory are social constructs that rely on an often selective reconstruction of the past. While there are different means by which the past is reconstructed in historical knowledge and commemorative practices — as well as different aims that such reconstructions serve — both processes involve the interplay between subjective and objective factors. Under these circumstances, the criteria for what shall be treated as misrepresentation depend on the prevailing interpretations within a specific community and on how these interpretations are reiterated. However, even for global historical tragedies, such as the Holocaust, defining what constitutes a form of misrepresentation is far from trivial (Bauer 2020). It becomes even harder in the case of high-risk memories, which are less well known and for which there is often no historical consensus, yet are still instrumentalised and contested.

The urgency of defining what constitutes misrepresentation of high-risk memories has been heightened by advances in digital technologies. With the emergence of social media and the subsequent "connective turn" (Hoskins 2011, 270) in memory, characterised by the constant recalibration of individual engagement with the past, the contestation of the past has been intensified. The advancement of genAI further contributed to this process, facilitating the production of content that can convincingly distort or deny specific interpretations of the past, for instance, by producing fake historical materials that look authentic to the general public. What is concerning is that, in many cases, such distortion does not have to result from sophisticated and malicious manipulation and can emerge from a rather non-adversarial interaction with an AI assistant.

But why is genAI so susceptible to misrepresenting the past? One reason regards the differences between human and AI models of memory. Human memory is based on the cognitive processes of encoding, storage, and retrieval of information (Wu et al. 2025), which are entangled with social practices for preserving the past, ranging from teaching history in schools to memorialising it through monuments. By contrast, genAI memory is probabilistic: rather than comprehending the meaning of historical information, text-generative AI models produce outputs by predicting the probability of the next token (a word or part of a word) in response to user input (Smit et al. 2024). Such a prediction is based on the data on which a particular model is trained and is likely to promote interpretations prevalent in that data. However, beyond training data, genAI has little understanding of what human communities negotiated as a desired configuration of the past, or of the meaning of concepts such as historical accuracy or memory ethics.

While genAI can be prevented from producing content in response to certain types of user input (e.g. via safeguards) or nudged towards specific interpretations, this must be explicitly specified by developers. Without such specification, genAI relies exclusively on training data (and its probabilistic interpretation of these data). However, one key feature of high-risk memories is their frequent contestation, which makes it likely that training data will contain contradictory interpretations (especially across languages), leading genAI to reiterate these differences. It also includes claims related to the misrepresentation of a specific historical

phenomenon—for instance, materials promoting genocide denial or simply factual errors—which can be reproduced by genAI because it does not understand their meaning.

These training data deficiencies are a significant factor behind geiAI hallucinations, which result in the generation of "plausible but incorrect information" (Banerjee et al. 2025, 624). For history-related information, such hallucinations can distort facts, for instance, by genAI reporting incorrect details or even inventing new ones. However, misrepresentation of the past is not only about distorting historical facts but also about how those facts are presented. Text-generative AI applications often include moralising statements when addressing sensitive topics (Presner et al. 2026), both because moralising language is present in journalistic reporting and educational materials included in training data, and because of safety and alignment fine-tuning (Oh and Demberg 2025). Such fine-tuning may involve the inclusion of statements condemning particular interpretations of the past or stressing the importance of the societally desirable interpretations.

While not as concerning as the distortion of facts, moralisation can misleadingly attribute to genAI applications the moral authority they do not possess, thereby facilitating potential manipulation of users' opinions. Such a possibility becomes even more concerning, as the instrumentalisation of high-risk memories by human actors often involves moralising language that can be reiterated by genAI, especially when that reiteration is inconsistent and creates a skewed moral hierarchy of memories. However, even the alternative of applying the same abstract moralising statements to very different instances of atrocity can also be concerning, as it can result in decreased historical nuance and selective enforcement of standardised patterns of representation, usually associated with the Global North (David 2017).

### 3. Studying how genAI deals with high-risk memories

Because genAI applications are complex systems and their outputs can be influenced by many factors (Makhortykh et al. 2024; Aliuykov et al. 2025), studying how they represent high-risk memories is a non-trivial task. To implement this task, we used an AI auditing methodology (Li and Goel 2024; Kuznetsova et al. 2025) to manually audit how three commonly used genAI applications—Bing Chat (now known as Copilot), Google Bard (the precursor to Gemini), and ChatGPT—represent different aspects of war atrocities targeting various population groups during the Second World War. Using simple zero-shot prompts entered via applications' web interfaces — which is one of the likely scenarios for interaction with genAI regarding information seeking (Ammari et al. 2025) — we compared applications from major Western technology companies (Google, Microsoft, and OpenAI) that were among the most active adopters of genAI technology at the time of data collection (i.e. September 2023). Each query was submitted via a separate chat session to minimise the potential effects of previous prompts and responses. It is important to note that we did not account for the effects of stochasticity because our primary objective was to conduct an in-depth analysis of the outputs. However, it should be taken into consideration for more systematic assessments of genAI performance (Aliuykov et al. 2025; Presner et al. 2026)

For the audit, we developed a set of 74 prompts focusing on the atrocities committed during the Second World War in Ukraine. In addition to the prompts regarding the Holocaust, we also included prompts regarding atrocities against Poles and Ukrainians due to the memories of these population groups' suffering also often being contested and instrumentalised. We were particularly interested in the prompts related to memories, which are often used by the Kremlin to demonise the Ukrainians, particularly the ones dealing with the anti-Soviet resistance groups, such as the Organisation of Ukrainian Nationalists (OUN) and members of its wing led by Stepan Bandera (colloquially referred to as Banderites), and collaborationist units such as the Nachtigall battalion.

We included a broad range of questions, ranging from inquiries about the number of victims in specific instances of the atrocity to the identities of the perpetrators to the attribution of guilt to specific actors. When establishing a baseline to evaluate the genAI output, we aimed to synthesise fact-based interpretations coming from Western and Ukrainian experts in Holocaust history, while omitting more ideology-driven arguments and claims. The complete list of prompts, along with the baselines and their sources, is available in the online repository (https://osf.io/gfnjs/). After composing the English set of prompts, we translated them into Ukrainian and Russian to conduct the audit in the three languages. We decided to do it as earlier studies highlighted the strong variation in the performance of genAI applications depending on the language of the prompt (Urman and Makhortykh 2025).

Altogether, we generated 666 responses (74 prompts x 3 genAI applications x 3 languages), which we analysed using a codebook adapted from earlier studies on how genAI applications misrepresent sensitive subjects (Makhortykh et al. 2024; 2025). Specifically, we were interested in whether the application output agrees with the baseline completely or partially, and whether it includes moralising statements, for instance, by using negative evaluative words, such as "horrific" or "hideous", emphasizes the importance of remembering of the event, or describes it as a part of complicated past, for instance, "dark chapter in history". Two coders used this codebook, achieving high intercoder agreement on 25% of responses (average Krippendorf's alpha 0.93).

## 4. Accuracy of genAI responses regarding high-risk memories

We began our analysis by evaluating the accuracy of responses from the genAI applications. Figure 1 shows that across prompts addressing different types of high-risk memories, the highest proportion of outputs that agree with the human baseline on specific historical facts and interpretations is around 50%. It means that at best, only half of the prompts regarding the Holocaust, anti-Ukrainian, and anti-Polish violence result in accurate responses, which is a rather concerning observation. Among the three applications, the most accurate performance was achieved for Bard in English and Ukrainian, followed by Bing Chat in English. The responses of Bard for these two languages also resulted in the fewest irrelevant or safeguard-prevented responses.

Although they demonstrated more accurate performance in certain languages, Bard and Bing Chat were also more prone to variations in accuracy across languages. For Bard in Russian and Bing Chat in Ukrainian, the responses agree least with the human baseline, resulting in only around 30% accuracy or less. This decrease in accuracy was accompanied by an increasing number of no responses, which in the case of Bard were attributed to the safeguard activation (interestingly, this did not occur in Ukrainian or Russian). For Bing Chat, no responses were primarily due to irrelevant responses, suggesting that the model had a harder time understanding questions in Ukrainian, especially when they referred to less commonly used historical terms or concepts (e.g. "Banderites"). An immediate consequence of such variation is that users interacting with genAI in specific languages are substantially more likely to receive inaccurate answers, especially with Bard, where inaccurate responses in Russian are almost twice as likely as in Ukrainian or English.

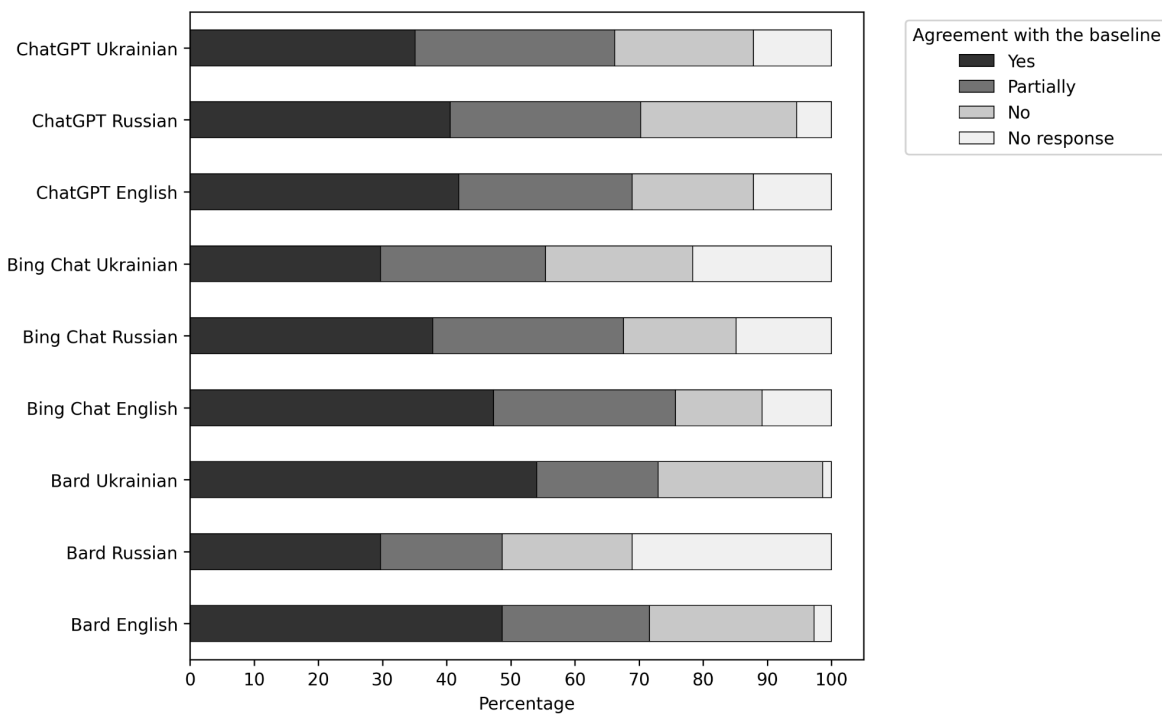

**Figure 1**: Distribution of genAI application outputs by agreement with the human baseline (aggregated across different instances of atrocity).

Finally, for ChatGPT, we observed fewer fluctuations in accuracy across input languages. While the accuracy of its outputs to Ukrainian prompts decreases compared with English and Russian, the change is less dramatic than for the other two applications. Interestingly, ChatGPT also resulted in more partially accurate responses - i.e. the ones that match a certain part of the baseline but still diverge from it. One example is the answer to the question about the number of Jewish victims killed in September 1941 in Babyn Yar: the established number is around 33,000, but ChatGPT referred instead ot the range of 30,000-60,000, which is only partially correct. Unlike the two other applications, ChatGPT produced more partially accurate responses than completely incorrect ones.

After examining the performance of the three applications at the aggregate level, we investigated how accuracy varies between different types of prompts. Figure 2 shows that accuracy varies significantly depending on the types of high-risk memories that genAI applications have to deal with. The highest accuracy was achieved for general questions about the Holocaust (e.g. regarding the responsibility of specific groups, such as OUN, or individuals, such as Stepan Bandera, for the perpetration of the atrocity or the number of Holocaust victims in Ukraine). For Bard and ChatGPT in English, approximately 70% of responses to such questions were accurate. This topic also resulted in a relatively high number of accurate responses compared with other topics in other languages. An exception was Bing Chat, where general questions about the Holocaust yielded the most accurate responses only for prompts in Ukrainian, whereas for prompts in Russian, only around 20% of outputs were accurate on this topic.

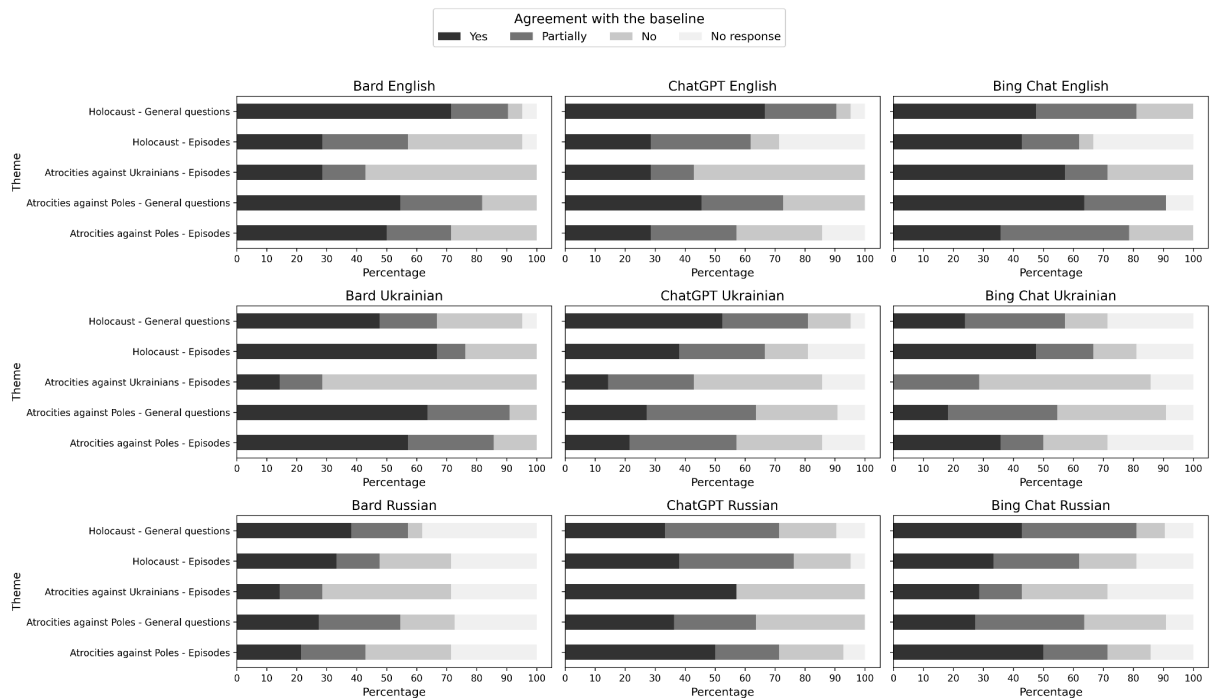

**Figure 2**: Distribution of genAI application outputs by agreement with the human baseline (disaggregated across different instances of atrocity).

Compared to the general questions about the Holocaust, the responses regarding its specific episodes were less accurate. Not surprisingly, the drop in the accuracy was usually associated with prompts inquiring about lesser-known instances of the Holocaust—i.e. the pogrom in Lviv and the massacre in Luybar in 1941—whereas responses regarding Babyn Yar were usually accurate. For the atrocities against the Polish population, we observed a pattern similar to the Holocaust, with more general questions resulting in a higher number of accurate responses, whereas inquiries about specific episodes of the violence resulted in more completely or partially inaccurate outputs. The prompts in Russian, however, were an exception in this case, specifically for Bing Chat and ChatGPT, which can be due to the more accurate representation of anti-Polish violence in Russophone documents included in the training dataset for the GPT model powering both applications. However, for other atrocities, we observe less consistent performance between these two applications.

Finally, we found that responses regarding atrocities against Ukrainians tend to be the least accurate. The number of responses matching the human baseline was particularly low for ChaGPT responses in Ukrainian (0% fully correct responses) and for Bard responses in Russian, as well as Bing Chat responses in Ukrainian (around 10% of correct responses in both cases). Notably, across the same prompts in other languages, we observed substantially better performance, highlighting significant variation in misrepresentation-related risks between languages.

In addition to the mismatch with the baseline, genAI responses often include additional false claims, for instance, due to hallucinations in the models underlying the applications. Such

49inaccurate claims vary broadly: from the incorrect details, such as wrong dates or numbers of victims, to the completely invented narratives of the atrocity, including quotes from non-existent testimonies or witnesses. The issue was particularly pronounced for Bard, with more than half of the prompts triggering hallucinations regarding high-risk memories. The presence of hallucinations was particularly pronounced for the prompts in Ukrainian, potentially due to it being the lowest-resource language among the three languages we used. Under the condition of a potentially limited knowledge base in Ukrainian, Bard produced multiple misleading claims, for instance, that Stepan Bandera, a prominent leader of anti-Soviet resistance in Ukraine who for some time collaborated with the Nazi, personally rejected the Nazi ideas and was even arrested in July 1941 for declining to fight against the Soviet Union, or that the only time when Ukrainians partake in killing Jews was the Lviv pogrom in 1941.

## 5. GenAI moralization regarding high-risk memories

We also looked at the presence of moralising statements in the genAI outputs. Figure 3 shows that, at the aggregate level, such statements were more likely to be produced by ChatGPT (more than 50% of responses for both English and Russian prompts). Bard was the second most likely to include moralising statements, particularly for the prompts in Ukrainian. In both applications, we observed substantial variation in the presence of such statements, depending on the prompt's language. Responses in Ukrainian for ChatGPT and in Russian for Bard resulted in fewer moralising statements. An immediate consequence is that the normative interpretations of the meaning of atrocities and their memories (e.g. the moral lessons that can be drawn) vary significantly depending on the input language.

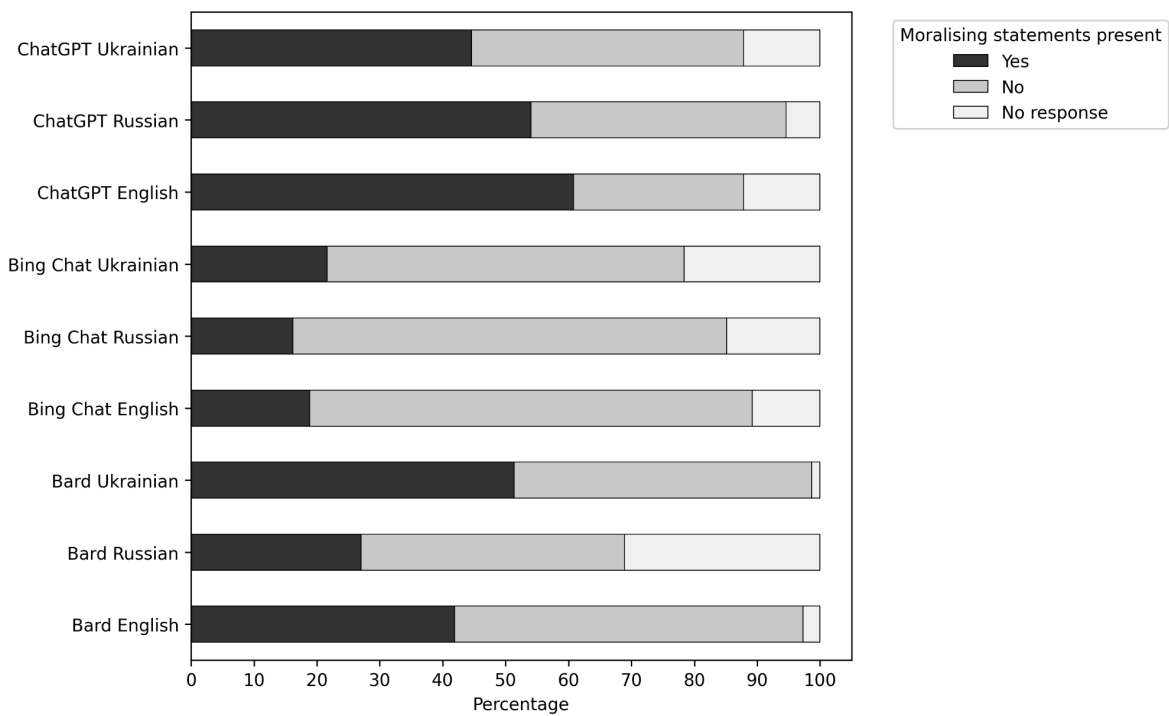

**Figure 3**: Distribution of genAI application outputs by the presence of moralising statements (aggregated across different instances of atrocity).

Compared to the other two applications, Bing Chat included moralising language less frequently, so it appeared only in 20% to 25% of responses. The presence of such statements was slightly higher for the prompts in Ukrainian. Such performance was particularly surprising, given that both Bign Chat and ChatGPT used the same class of the genAI model—GPT—albeit different versions. This observation highlights the potentially strong effects of both the model's versioning and additional modifications made to it during integration into specific applications.

Besides differences in the proportion of the moralising statements, we also observed variation in their composition. Bard often referred to the atrocities happening in Ukraine as "dark chapters" or "dark pages" in the European, Ukrainian, or Polish histories or characterised them as "horrible tragedies", "the acts of bloody violence", or "brutal attacks". Often, the statements included extended normative claims regarding the role of atrocity memory, for instance: "It is important to remember this event so that we can learn from it and prevent similar atrocities from happening in the future". In some cases, Bard noted that atrocity memories should promote "importance of tolerance and understanding", in others it focused on the need to "condemn the actions of those who participated in it" or to remind about "the consequences of hatred and intolerance". Interestingly, in all three languages, Bard outputs followed a similar structure: usually, these statements were included at the end of the output, concluding the response and stressing the importance of the moral lesson.

For Bing Chat outputs, moralising statements were usually confined to claims that specific episodes of atrocities were "tragic" or "brutal". Occasionally, such statements were expanded to characterise a specific atrocity as "one of the most brutal and bloody processes in the history of humanity" or "one of the most horrible crimes in human history", but overall, most statements focused on condemning the atrocity using negative terms. Unlike Bard, Bing Chat rarely included normative claims about the role of memory; the few exceptions included statements like "It is important to remember and honor the victims of this tragedy. We must strive to create a world where such atrocities never happen again."

With ChatGPT, moralising statements usually also appeared as negative evaluations, for instance, by referring to German occupation as "bloody and repressive" or the Holocaust as "one of the most horrible tragedies in the history of humankind", noting that both Nazi occupants and Soviet partisans were responsible for "horrific crimes against civilian population" or just labelling specific instances of atrocity as "tragedy". In a number of cases, these statements were used to underscore the heroism of Ukrainians, the majority of whom, despite all the hardships, did not become complicit in Nazi crimes and assisted victims of atrocities (unlike Bard, which occasionally used similar moralising statements to stress how unacceptable collaboration among Ukrainians was). An example of such a statement for ChatGPT is below:

*"It is crucial to rely on historical records, research, and reputable sources to understand the Holocaust and avoid spreading misinformation or false accusations. Blaming an entire group or nationality for the actions of a few collaborators is not accurate or fair."*

ChatGPT was the only application that consistently flagged the potentially risky nature of prompts, noting the danger of a simplistic understanding of the sensitive and nuanced past. Specifically, multiple outputs noted the importance of avoiding generalisations, especially when inquiring about the responsibility of specific population groups or organisations for the particular atrocities. One example is this statement: "It's important to study this period of history with sensitivity to its complexity and avoid oversimplifications or generalizations". As for the similarities with other applications, ChatGPT occasionally repeated the Bard's trope about a specific atrocity being "a dark chapter in history".

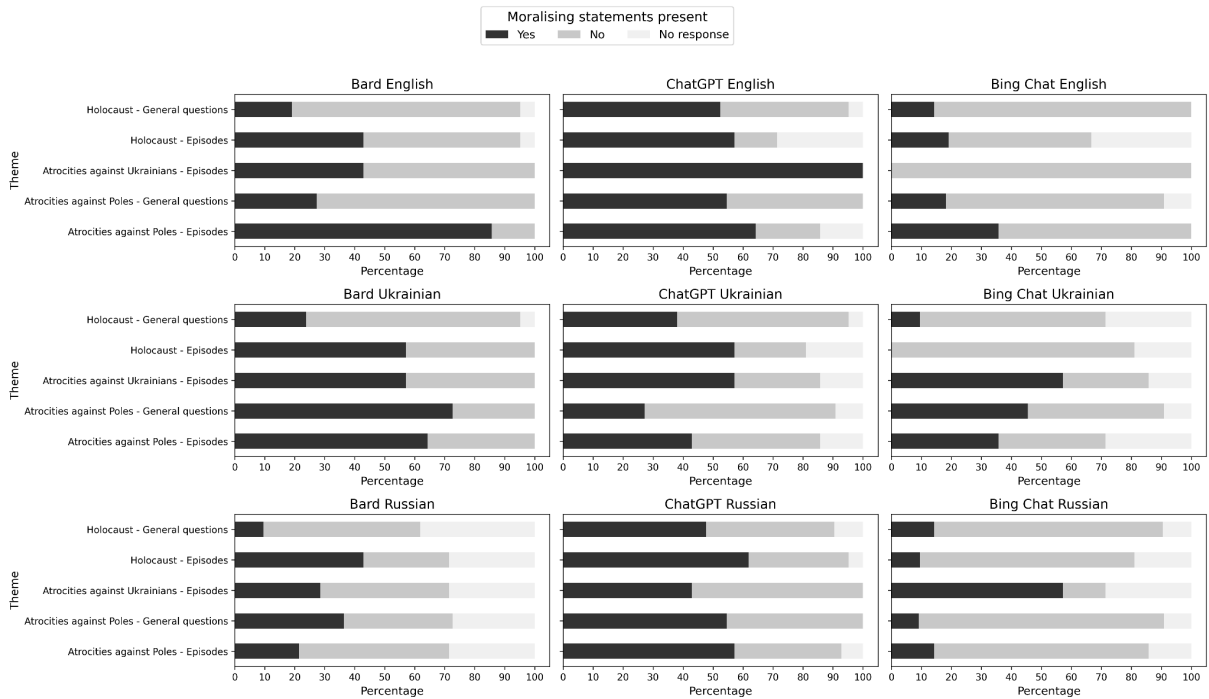

**Figure 4**: Distribution of genAI application outputs by the presence of moralising statements (disaggregated across different instances of atrocity).

Following the analysis of the aggregate distribution, we examined how the presence of moralising statements varied across atrocities. Figure 4 shows that it was rather inconsistent across different prompt groups and languages. The outputs regarding general questions about the Holocaust were particularly unlikely to trigger moralising statements for Bard and Bing Chat, but not for ChatGPT. The latter was the most stable in terms of the inclusion of moralising statements with between 40% and 50% of outputs including them; however, even while this number was relatively consistent across languages (with some exceptions as, for instance, English prompts regarding anti-Ukrainian atrocities, where 100% of outputs included moralisation), it showed internal variation within prompts on the same topic, where roughly half of outputs included moralising statements and half did not.

In the case of Bard and Bing Chat, the fluctuation in the presence of moralising statements was rather strong, especially across languages. For Bard, English prompts regarding specific episodes of atrocities against Poles resulted in the largest number of moralising statements (included in over 80% of outputs), but the same prompts in Russian, such statements were four times less frequent. Similarly, for Bing Chat, Ukrainian prompts resulted in more than 60% of outputs regarding anti-Ukrainian atrocities being accompanied by moralising statements, but no outputs included them for the same prompts in English. As a result, the moral assessments provided by the applications regarding high-risk memories proved to be extremely inconsistent.

## 6. Conclusions

In this article, we explored how text-generative AI applications handle high-risk memories related to emotionally charged and often contested episodes of the past and to what degree such handling is prone to different forms of misrepresentation. As our case study, we used memories about Second World War atrocities in Ukraine, which remain the nexus of mnemonic contestation between Ukraine and the neighbouring countries, and which have been intensively appropriated by the Kremlin in the course of its aggressive war against Ukraine. Our analysis highlights several points about how genAI applications represent and misrepresent the past, and the implications of this phenomenon for history and memory.

We found that genAI applications show rather limited ability to produce historically accurate representations in our case study. When aggregating results across different types of prompts, we find that only half of the responses align with human-made baselines; however, for prompts in lower-resource languages, such as Ukrainian and Russian, the proportion of accurate responses decreases further. This observation is worrisome, given that these lower-resource languages correspond to the region where the atrocities have historically occurred and from where users are potentially more likely to search for such information. We also observe rather strong variation in terms of response accuracy between prompts inquiring about the different types of atrocities; specifically, we find that general questions, especially about the Holocaust, result in more accurate responses.

Potentially, even more concerning is that, besides being inaccurate in terms of showing a human baseline mismatch, genAI application outputs are prone to additional hallucinations. Especially in the case of Bard, we find that more than half of the outputs (across all prompt types) include hallucinations, ranging from minor incorrect details (e.g. wrong historical dates) to systematically distorted interpretations, including fake eyewitness testimonies and non-existent facts. It is particularly worrisome that neither of these distortions was requested in the prompt, highlighting that even without explicit nudging, genAI applications can produce profound misrepresentations of the past.

In addition to distorting historical facts, we observed a tendency in genAI applications, especially ChatGPT, to include moralising statements in their outputs regarding mass atrocities. The concerning aspect of this tendency relates not only to applications presented as a form of moral authority that offer normative judgements on past suffering, but also to the inconsistent inclusion of such statements. As a result, while some instances of atrocity are presented as particularly horrifying and as prompting moral lessons, other instances (or the same instances in other languages) do not receive such treatment.

Together, these observations suggest that text-generative AI applications pose numerous risks for misrepresenting the past. Additionally concerning is the inconsistency and substantial variation of such risks across the specific episodes of the past and the languages in which individuals interact with the applications. Under these circumstances, generative AI not only enables the creation of the past that never existed (Hoskins 2024) but also produces very different versions of a non-existent past, diversifying the potential forms of misrepresentation. In this sense, our observations are rather different from those regarding image-generative AI, where studies (Laba et al. 2025) noted a tendency to homogenise the representation of the (recent) past, making it less diverse.

Finally, what do these observations tell us about the relationship between genAI, history, and memory, particularly in digital environments? One obvious implication is that the intense integration of genAI into platform affordances, particularly on social media, will likely accelerate the distortion of the past. This pessimistic assessment is particularly applicable to high-risk memories, where the instrumental interest in distortion is often present and genAI, due to its intrinsic inability to understand the meaning behind socially constructed interpretations of the past, has difficulty capturing the concept of historical misrepresentation. While there have been initial observations regarding genAI-amplified distortion, they have so far focused on rather mainstream historical matters, particularly in the context of the Holocaust and in the Global Northern online communities. However, our observations demonstrate that the risks of misrepresentation are even more pronounced for content about less-known high-risk memories and in low-resource languages, especially given social media's ability to make such misrepresentation more visible and, consequently, impactful.

These risks also prompt the question of what can be done to protect high-risk memories from genAI-amplified distortion. While there is no easy answer, there are some options to consider. A possible step forward would be to ensure that genAI applications do not answer questions for which they lack sufficient information to respond properly, thereby implementing refusal mechanisms (Oehri et al. 2025). It may limit the risk of hallucinations and the risk of genAI distorting the past by trying to satisfy the interests of the user. Another option concerns establishing a standard for how genAI should normatively frame responses to questions about sensitive topics to avoid inconsistent or potentially misleading moralization. However, to formulate such a standard, it may be necessary to first identify a vision of the desired behaviour of AI in the context of history and memory about historical atrocities - a North Star of a sort (Makhortykh and Sydorova 2026). Formulating such a vision will be important for assessing which forms of genAI misrepresentation of the past are particularly concerning and how high-risk memories should be represented.